# Physics at the Transition between Bounded and Unbounded Trajectories

Alasdair Macleod
University of the Highlands and Islands
Lews Castle College
Stornoway
Isle of Lewis
UK
Alasdair.Macleod@lews.uhi.ac.uk

*Abstract:-* The electromagnetic interaction is characterised by discrete states for bound systems in contrast to continuous states for unbound systems. The difference merely arises because the characteristic equations do not exhibit the same behaviour for negative and positive energy parameter values, thus the apparent distinction between bound and unbound states can be considered purely descriptive and completely superficial – there is no indication that bounded and unbounded systems are subject to differing physical laws. However, the remarkable suggestion has been made that there exists a behavioural distinction between bound and unbound states of systems under gravitational influence. This notion is critically evaluated here. At the very basic level, a severe problem is found in defining a local concept of boundedness consistent with the current understanding of the gravitational interaction. Nevertheless it is difficult to exclude the possibility that bound and unbound systems are dynamically distinct, a distinction that may be relevant to existing cosmological and astronomical anomalies.

**I Introduction**

Dynamic systems may be categorised as bound or unbound; if the sum of the kinetic and binding energies is less than zero, interacting entities are considered bound; and if greater than zero, unbound. The state of a system where the total is zero is a matter for debate. One can interpret the terms 'bound' and 'unbound' either as an indication of the level of constraint, or, equivalently, a description of the spatial extent of trajectories.

In this paper, we will consider whether it is probable that bound and unbound gravitational systems should be subject to fundamentally different forces. On the face of it, this is a strange proposal quite at odds with the current understanding of gravitation because it implies a distinction between how General Relativity (GR) is applied locally and globally - such a division is incompatible with GR. After all, GR is currently applied effectively to systems ranging from orbiting global positioning satellites on the small scale to the prediction of the rate of expansion of the Universe on the grand scale. Also, the mathematical representation of the physical Universe is particularly effective and does not admit specific terms sensitive to boundedness, hence, in dynamic terms, the transition between bound and unbound is predicted to be uneventful and featureless. Of course mathematical representations are associated with models and the requirement that models should be scale independent with no difference on the very small and very large scales is an *a priori* assumption that may not be justified. Indeed, because the property of 'boundedness' is absolute and non-transformable, it would be technically possible to define two independently consistent variants of GR separately applicable to bound and unbound systems. The penalty would be the presence of discontinuities at the interface. Physically, this would mean anomalous observable effects at the transition between bounded and free trajectories, but it may be that these really do exist and map onto spacecraft trajectory irregularities that have already been recorded (specifically the Pioneer[1] and Flyby[2] anomalies).

In fact, the motivation for this investigation is some apparently innocuous but in fact extraordinary statements that have been made in this context, particularly concerning the Pioneer Anomaly. For example [2, 3, 4]

> 'The Pioneer anomaly was found on the spacecraft following hyperbolic, unbound escape trajectories… There also exist anomalies seen in hyperbolic planetary flybys. This all emphasizes how the transition from bound to escape orbits has never been well characterised.',

or

> 'It may be that the modification [*the MOND acceleration*] enters the Pioneers motion, which corresponds to unbound, hyperbolic motions, and the motion of bound, and quasi-circular trajectories in a different way.',

and

> '.. the unbound hyperbolic trajectory would have an additional acceleration of size of the Pioneer anomaly. This would be confirmed if the Pioneer anomaly turned on only at Saturn flyby for Pioneer 11 and only at Jupiter flyby for Pioneer 10.'.

Whether or not intentionally, these statements seriously challenge established physical models and suggest a failure of the current paradigm, implying the need for a significant revision; though this presents difficult philosophical issues. No existing ontology admits a distinction between bound and unbound systems and, in particular, it can be shown that such a concept is irreconcilable with a substantive interpretation of spacetime.

However it would be negligent to dismiss the idea out of hand purely through philosophical considerations. Sources that are unbound with respect to the Earth-based observer have been extensively studied: the cosmic background radiation; distant supernovae; the Pioneer probes. In all cases, there are anomalous features that hint at the need for some modification to our world model. While the discrepancies are small in relative terms, this does not guarantee that the associated changes in the model needed to absorb the anomalies are minor. There have been many new proposals presented to explain away the anomalies, including the proposition that the form of the gravitational force changes over distance (the MOND proposal makes this suggestion, for one[3]). In fact many of the ideas seriously undermine GR and have a complexity and contrivance that are aesthetically unappealing. Though outlandish and counter-intuitive, it is possibly less destructive to the existing knowledgebase to identify the anomalies with some fundamental behavioural difference exhibited by bound and unbound systems and influencing the dynamics in the way hinted at by the authors of the quotations above.

One possible behavioural factor is that retardation effects differ between bound and unbound systems, but this attractive proposal does not stand up to scrutiny: Particles do not act instantaneously on another over distance; the influence is subject to a propagation delay proportional to the propagation speed of the mediating quanta. This generally gives rise to retardation effects, but these are limited because the equations describing the force incorporate an extrapolation effect into the solution functions that determine the dynamics[5]. Charged particles appear to extrapolate the velocity of a charged source and respond accordingly, whilst masses subject to the gravitational force extrapolate both the velocity and acceleration of the source. Consequently, charged particles must be accelerated and gravitational systems must be subject to a change in acceleration before retardation effects appear. The procedure is of interest here because a bound charge is no longer subject to the classical dynamic equations and becomes inaccessible to the observer – the charge no longer continuously radiates although we can infer it is still subject to an internal acceleration. This curious anomaly is quantum in origin. Could something similar exist with the gravitational attraction though the quantum description is not yet known? The answer is no – the Pioneer spacecraft are certainly unbound, but their trajectories do not show the characteristic signature of retardation effects in either distance or velocity[1]. In a sense, this is simply more evidence for the correctness of GR and how it is *not* apparently scale dependent.

However, retardation is not the only possibility. In this paper we consider in some detail the alternative idea that the transition from a bound to an unbound trajectory is characterised by an entity joining into the Hubble flow.

**II Flyby Anomalies**

The flyby anomalies are small but unexpected step increases in velocity experienced by several spacecraft using the Earth for gravity assist manoeuvres. The manoeuvre is basically a slingshot technique that rotates the spacecraft velocity vector with respect to the Earth, but results in a nett increase (or decrease) in velocity (and kinetic energy) with respect to the barycentre once the orbital velocity of the Earth is factored in.

The anomalous effect is variable in magnitude and is not present in all Earth gravitational assist events (EGAs). The results of a number of events is summarised in Table 1. The Gallileo and NEAR events are discussed in detail by Antreasian and Guinn[6] (AG), and the dynamics of all the encounters listed in Table 1 are presented by Anderson, Campbell and Nieto[7] (ACN).

**Table 1** Flyby anomaly for the EGA events analysed by ACN[7]. $v_\infty$ and $v_F$ are respectively the hyperbolic excess velocity and the velocity at perigee. Remarkably, the anomalous effect appears not to be correlated to orbital parameters such as eccentricity hence these figures are not presented. In a positive assist, the barycentric velocity increases, and decreases with a negative assist.

| Spacecraft | Type of Assist | $v_\infty$ (km s$^{-1}$) | $v_F$ (km s$^{-1}$) | $\Delta v_\infty$ (mm s$^{-1}$) | $\Delta E$ per kg (J kg$^{-1}$) |
|---|---|---|---|---|---|
| Galileo | Positive | 8.949 | 13.738 | 3.92±0.08 | 35.1±0.7 |
| NEAR | Negative | 6.851 | 12.739 | 13.46±0.13 | 92.2±0.9 |
| Cassini | Positive | 16.01 | 19.03 | ~0 | ~0 |
| Rosetta | Positive | 3.863 | 10.517 | 1.82±0.05 | 7.03±0.19 |
| Messenger | Negative | 4.056 | 10.389 | ~0 | ~0 |

The hypothesis that the anomaly is associated with the transition between bounded and unbounded paths can be tested using the available data. On the face of it, this seems nonsensical because the spacecraft always remain bound to the Solar System through the encounters, and in the reference frame of the Earth they occupy hyperbolic unbounded orbits – there is no evidence of any sort of transition either from the point of view of the Sun or the Earth: The Solar escape velocity along the Earth's orbital path is 42.1 km/s and the orbital velocity of the Earth is 29.8 km/s (mean). Because of vector alignment issues and the Earth field, ACN show the spacecraft never make up the difference. However, progress can be made if we relax the definition of boundedness from that described in the first paragraph of this paper and permit an arbitrary energy level to be associated with the transition point, more or less constant for a particular system under scrutiny (we will see later that there may be some justification for this because it is uncertain whether spacecraft share our apparently clear understanding of boundedness).

Let $S^0$ and $S^1$ describe the bound and unbound states of the spacecraft respectively within the Solar System. Postulate the existence of an energy associated with these states (distinct from potential and kinetic energy) with $E(S^0) < E(S^1)$. This can be considered analogous to a phase shift in matter when the energy difference can be likened to latent heat. If an entity moves from an unbound to bound state ($S^1 \rightarrow S^0$), energy is released to appear as an addition to the normal kinetic energy. If an entity moves from a bound to unbound state ($S^0 \rightarrow S^1$), the same quantity of energy is absorbed from the surroundings. An entity hopping between states from bound to unbound and back to bound ($S^0 \rightarrow S^1 \rightarrow S^0$) may emerge with a nett energy gain if they are in the process of interacting with other bodies. These other bodies then share in the energy exchange.

ACN have generated graphs showing how the total barycentric energy per unit mass changes about the point of closest approach. It is proposed here that the value $-2.75\pm0.05 \times 10^8$ J kg$^{-1}$ should be identified with the transition between $S^0$ and $S^1$. An energy per unit mass that is *more* negative represents, $S^0$, the bound state. Energy per unit mass that is *less* negative than the transition value indicates, $S^1$, the unbound state. It is possible to determine the states of each of the spacecraft through the EGAs for this threshold value by an examination of the data in ACN (Fig. 1).

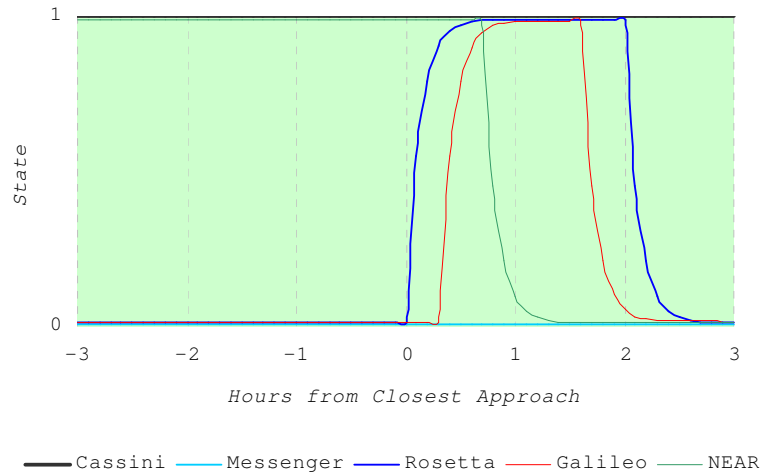

**Figure 1** The state of each of the spacecraft during the EGA events assuming a barycentric total energy density of $-2.75\pm0.05 \times 10^8$ J kg$^{-1}$ represents the transition between $S^0$ and $S^1$. Although the transition is described as a step change, this is clearly unphysical. The transitions are likened to a mass $M$ (with velocity $v_M$) dropped onto a moving conveyor belt (velocity $v_c$) – it takes the frictional force time to bring the relative speed of the mass and the belt to zero (using constant $k$ to describe the frictional coefficient). The characteristic equation is $dv_m/dt=k/m(v_c-v_M)$ with solution $v_M=v_c(1-e^{-kt/M})$. The transitions in the graph are modelled in this fashion using arbitrary parameter values.

The Cassini and Messenger events do not involve a change of state, hence there is no anomalous increase predicted. The NEAR event is a change from an unbound to a bound state with a gain in free energy because $E(S^1) > E(S^0)$. The Rosetta and Galileo events involve a transition from bound, briefly to unbound and then back to bound again. Energy is lost and then gained with a residual that is much smaller than the NEAR event.

There is further supporting evidence for the hypothesis from the profile of the Galileo and NEAR range residuals presented by AG and reproduced below in Fig. 2. It is clear that the transitions are not truly instantaneous, but have a finite rise time. The Galileo event on the left shows evidence at the transition region of a much higher initial rise then a quick fall to the final value, exactly as expected if the spacecraft briefly entered the $S^1$ state. The pulse width is of the order of an hour just after closest approach (in agreement with Fig. 1). The unusual diurnal oscillation in the NEAR data is clear evidence of a change of physical behaviour before and after the event (and this is the only one of the flybys where an overall state change occurs). However, as will be seen later, the oscillation is expected to be associated with the unbound state, not the bound state as appears to be shown by the residuals – but it may be that the inverted appearance of the signal is the result of an overzealous removal of the true effect before closest approach by over applying Earth-model corrections.

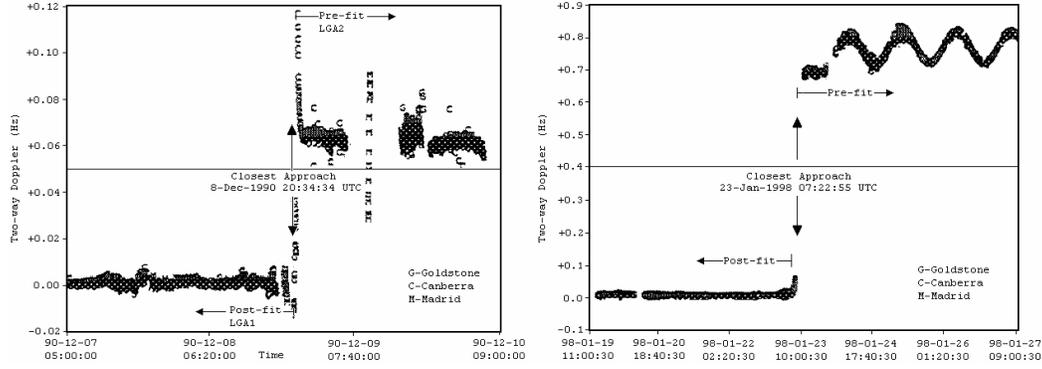

**Figure 2** Range residuals for the Galileo (left) and NEAR (right) EGA events (from AG, with a minor clean-up to the scale and labelling to improve clarity).

There seems to be qualitative evidence that the flyby anomalies are associated with either a real or apparent state change with respect to the Earth-based observer. We have suggested that the states are associated with bounded and unbounded trajectories. In that case, what is the significance of the non-zero value of the transition energy level, and if we are to associate different physics with bound and unbound systems, how is the state of boundedness known to the entity itself so that it can dynamically respond in the appropriate way?

**III A Local Definition of Boundedness**

If we adopt the field interpretation of Newtonian gravity, it is, in principle at least, straightforward to determine whether or not a body is gravitationally bound. The gravitational acceleration can be expressed in terms of a potential $\Phi$, where the dynamical response then takes the normal form $\partial^2 x^i / \partial t^2 = \partial \Phi / \partial x^i$. Because gravity is conservative, the total binding energy at any position is easily calculated by the scalar addition of the potential with respect to all gravitating mass and this of course is incorporated into $\Phi$, but it must be emphasised $\Phi$ is not an observable; only the gradient is. If the kinetic energy exceeds the potential, the object is unbound. However, it is known the field model of gravitation is incorrect (because it is impossible to formulate a scalar, vector or tensor description for $\Phi$ that is invariant without recovering the equations of GR) but there are difficulties in translating the concept of binding energy to GR where gravitation is understood to be the response of an object to the local spacetime curvature represented by the Einstein tensor $G_\mu^{\ \nu}$. The curvature represents the combined effect of all gravitating mass (including electromagnetic fields) as specified by the stress-energy tensor $T_\mu^{\ \nu}$. The binding energy is certainly coded into the Einstein tensor, but cannot be extracted from the spacetime manifold merely by measuring the local curvature. For instance, there are locations in the Solar System where the curvature and hence the acceleration is zero because of the favourable opposition of gravitating bodies, but any object located there certainly cannot be considered free by the common definition. At the very least, it is necessary to reference against a flat Minkowski metric, and, in any case, it is difficult to understand how an object blindly following the geodesic path through spacetime should be expected to execute this calculation, or any calculation.

An alternative approach is to refer to the spatial extent of a trajectory as the test for boundedness. If an object is on a path that permits an escape to infinity, it is unbounded. Of course, future events such as collisions affect the true state of boundedness by this definition, so knowledge of the future would be required (unless the Universe happens to completely deterministic). This is of course completely acausal, introducing nonlocality and a completely unjustifiable set of new problems.

The situation is even worse when we admit the possibility that the anomalous increase in velocity is *relative* to the Earth rather than an absolute (at least with respect to the Solar System). Certainly the spacecraft radial velocity is known to the observer, but there is no way a relative observer can know the depth of the potential well at the position where the observed spacecraft happen to be, hence it is truly unfeasible to determine boundedness from afar. In addition, if it really is a relative effect, it is necessary to make a distinction between unbound with respect to the Solar System or the Galaxy or even bound Galactic clusters, and adjust the dynamics accordingly – how many separately expanding spacetimes does this require? On the other hand if it is an absolute effect and not an observational anomaly involving the Earth or its dynamics then all Solar System bodies are subject to the strange energy threshold. Then any body that is more than 3 AU away from the Sun will have a total energy above the threshold and must be considered unbound.

This extreme confusion is why the apparently innocuous statements made concerning the Pioneer anomaly and quoted earlier are so difficult to incorporate into our current models – it is impossible to see how an object should know if it (or anything else) is bound, either in an absolute or relative sense. One would have to introduce a new term into GR representing a direct interaction with $\Phi$. There is one dubious bonus though: Because it is so difficult to have a local definition of boundedness, the threshold value that emerged from the previous analysis, $-2.75\pm0.05 \times 10^8$ J kg$^{-1}$, is really no stranger than the value of 0 one would intuitively expect – neither should be significant so both are equally unlikely. Regardless of whether or not there is an association with boundedness, the appearance of this quantity, which seems to unify the response of the spacecraft through EGAs, is extremely curious. What is the implication of this value?

**IV The Pioneer Anomaly**

Another puzzling question is the origin of the energy difference supposedly associated with the states $S^0$ and $S^1$? The NEAR event is the only one that is postulated to lead to a permanent change of state, and one would presume the increase in energy registered for that event is characteristic of the true energy difference between the two states. The actual value (per unit mass) of $92.2\pm0.9$ J kg$^{-1}$ is suggestive of an association with the Hubble expansion, and encourages the hypothesis that the transition from bound to unbound state is connected with a body joining the Hubble flow (leaving open for now the question about the threshold energy condition and whether the transition effect is absolute or relative). The analysis that follows is based on assumptions we will come back to critically assess later, but the bare argument proceeds roughly as follows. At the position of the Earth, the Hubble velocity of expansion with respect to the barycentre, $v_H$, is $rH_o$, where $r$ is the distance to the Sun, 1 AU. We can construct many dimensionally valid expressions for an energy associated with the expansion that differs from, and is in addition to, the customary kinetic energy of the expansion velocity: The simplest form is $mv_Hc$ which gives an energy per unit mass of $102\pm10$ J kg$^{-1}$, acceptably close to the actual value.

This definition is tentative of course, but making the energy difference $E(S^1)-E(S^0)$ a function of position is not without consequences, and implies further unexpected (to our theories as they currently stand) effects if the theory is correct. The detection of these effects would then lend considerable support to the general correctness of the expression chosen for the energy difference between states. The main problem that needs addressing is that the definition of energy as $mrH_oc$ permits a clear violation of energy conservation through the execution of any cycle like the one shown in Fig. 3. Limitless energy can seemingly be created out of nothing.

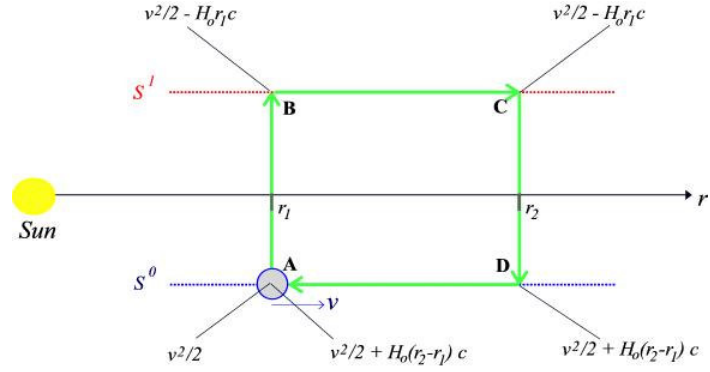

**Figure 3** Assuming $H_o$ can be considered constant, and ignoring the work done against the gravity (which is always recovered) the cycle A→B→C→D→A shown generates a quantity of energy per unit mass of $H_o c (r_2-r_1)$ on every circuit.

It is easy to see that the fundamental cause of the energy violation is the change in recession velocity accompanying proper (or peculiar) radial motion. A balancing force taking the general form $F^i = -\partial E/\partial x^i$ is needed, but will clearly be strictly radial because the energy has been defined purely a function of $r$ (for the moment at least). Let an unbound body at position $r$ from an observer at the origin move with peculiar velocity $v_P$. By the previous definition, the body is subject to an acceleration of $-1/m\, \partial/\partial r (mrc/T_o)\hat{r}$ which can be conveniently labelled $a_o$. Note that we let $H_o$ equal $1/T_o$, the current age of the Universe. Differentiating and identifying $\partial r/\partial T$ with the sum of the recession velocity and the radial component of the peculiar velocity, we get:

$$\boldsymbol{a_o} = -\frac{H_o c}{\left(1 + \dfrac{r^2 H_o}{\boldsymbol{v_P}\cdot\boldsymbol{r}}\right)} \hat{r}. \tag{1}$$

This acceleration is always directed towards the origin (except where the proper and recession velocities are about the same magnitude) and actually represents a *deceleration* for objects with peculiar velocity in the direction away from the origin, and ensures energy conservation is recovered: Referring back to Fig. 3, work is now done through section $\overrightarrow{BC}$ of the cycle that cancels the energy formerly gained. If we run the cycle in reverse, energy is now gained moving along $\overrightarrow{CB}$ because of the acceleration in that direction, again conserving total energy.

Equation (1) is very much simplified if the modulus of the radial motion far exceeds the modulus of the recession velocity:

$$\boldsymbol{a_o} = -H_o c\, \hat{r}. \tag{2}$$

The acceleration magnitude is constant and is $\approx -6.95 \pm 0.55 \times 10^{-10}$ m s$^{-2}$ using the currently accepted value of $H_o$ and its error margin. Unbound objects moving away from the Solar System will satisfy this velocity condition and must experience a deceleration of that magnitude if the original definition of the energy associated with a transition between bound and unbound stated is correct. Unfortunately there are few cases of objects leaving from the Solar System for which precision velocity and/or range data is available. The best examples are the Pioneer 10 and 11 probes that achieved escape velocity in 1973 and 1979 respectively. Path information from 1980 until contact was lost certainly indicates an unmodeled deceleration force affecting both spacecraft. An extensive analysis[1] concluded there was a constant deceleration affecting both craft of magnitude 8.74±1.33 x $10^{-10}$ m $s^{-2}$. Whilst it is tempting to identify this with the deceleration predicted by equation (2), there is a significant discrepancy – the error bars do not meet. However, there is an indication that the observed deceleration effect is not simply a constant; there are clear annual and diurnal overtones that may be related to the motion of the Earth relative to the barycentre, or are a residual effect of the movement of the observer relative to the spacecraft in excess of the normal Doppler corrections which are routinely applied. It may be that equation (2) simply predicts the constant portion of the total deceleration and is therefore correct. The oscillatory effects need to be examined to validate this conclusion, but there are more fundamental issues that need to be considered first: We have assumed throughout the anomalous effects are always directed towards the centre of mass of the system in which the observer is bound. We consider the justification for this in the next section.

One point to note is that there is something distinctly unphysical about equation (1). If the proper motion of the observer is exactly opposite to the expansion velocity, the acceleration becomes infinite. For example, if the anomalous Pioneer acceleration were directed towards the Earth rather than the barycentre (which appears not to be the case), a huge acceleration spike would be registered about a month after the sun occults the spacecraft (because the velocity of the Earth with respect to the probes ranges from –17 to +41 km $s^{-1}$ over a period of revolution and must pass through the discontinuity). For the Pioneer 10 probe, this is around the middle of July each year. The presence (or absence) of this spike unfortunately cannot be used in an effort to find the true direction of the anomalous acceleration as the data is always noisy around that time because the signal is affected by passage through the Solar corona, and in any case, the signature would be extremely transient, lasting only a few seconds. However, the prediction of an infinity, no matter how transient, is an indication that the chosen energy definition, $mrH_oc$, is not completely correctly and needs to be refined as more detailed data becomes available.

We may conclude that there appears to be reasonable evidence supporting the hypothesis that a quantity of energy of the order of $mv_Hc$ is released when a mass moves from an unbound state. If this energy really exists, what is it and where does it come from? It is hard to do anything other than speculate, but it may be significant that as we look back in distance and time towards the origin of the Universe to the point where the recessional velocity reaches the speed of light, the spontaneous creation of matter into a bound state becomes energetically favourable.

**V The Local Cosmological Expansion**

Associating the transition from a bounded to unbounded orbit with an object joining the Hubble flow is counter to the commonly held view that the cosmological expansion does not proceed in the local galactic neighbourhood. But is there any real evidence for this common view? There are only theoretical arguments, the most significant possibly being that proposed by Cooperstock *et al.*[8] in 1998. It had formerly been argued that the expansion only occurred outside the galaxy[9] but that implied the existence of a 'switch-on' point, and there was no real theoretical justification for such a transition, or any insight into the physics at the transition where an object joined the Hubble flow. Cooperstock *et al.* circumvented the problem and established the current stance by claiming the expansion is universal, acting on all scales; but that the effect is undetectable in our immediate surroundings because of the restraint applied by the dominant forces overwhelms the miniscule motion of the expansion. This argument is now widely accepted, not least because it clears the potentially awkward problem of a transition point. However, it is hard to agree with the argument presented to support the case

The paper investigated the transition between the co-moving and observer frames. In general relativity, cosmological observations are presumed to be made from a local inertial frame (LIF), an idealised non-rotating, non-accelerating frame in the vicinity of the Earth. The dynamics of the Universe is described by the time-dependent scale factor in the Friedman-Robertson-Walker metric and Cooperstock *et al.* argue that the LIF is an instantaneous tangential approximation to the curved spacetime of GR. The assumption is made that the LIF is exempt from the expansion by its very nature. The basis of their argument is outlined in the following two excerpts:

> 'Although the cosmological expansion is described by the time dependent scale factor in the FRW metric, and we believe affects lengths at all scales, the curved spacetime manifold can be locally approximated by its (flat) tangent space at every spacetime point $p$ . … This frame is the one in which astronomical observations are carried out. Thus, the effect of the cosmological expansion is seen to be negligible locally and grows in significance with distance, reaching full import on the cosmological scale.'

and

> 'It [the local inertial frame] is the locally inertial frame based on a geodesic observer and it continues to be locally inertial, following the observer in time.'.

In fact, it is probable this approximation and indeed the entire methodology are incorrect. Expressed in co-moving coordinates, the local reference frame is clearly subject to the expansion as well, whereupon no expansion can be perceived in the co-moving frame. The argument seems to require the LIF be described in coordinates that are independent of the expansion. In this model there is only one manifold, and it is expanding. How are the constant coordinates to be maintained? In effect, the approximation procedure adds in the non-expansion of the local reference as a rather subtle assumption from the outset rather than deriving it. The correct approach would be to operate entirely in co-moving coordinates and show that the LIF is contracting over time. This is completely equivalent to the local frame not being subject to the cosmological expansion. It is therefore concluded that Cooperstock *et al*. have not demonstrated the expansion operates within the Solar System. In effect, a non-expanding reference frame that forces may use to detect the expansion does not exist; forces are expressed in terms of the position coordinates defined in expanding spacetime and are therefore unaware of the expansion.

By denying the argument of Cooperstock *et al*., we are introducing two coexisting spacetimes, but both are not accessible at the same time. An object has access to one or other depending on its state of boundedness. The movement between bound and unbound state is then a change in reference frame. The apparent distinction between bounded and unbounded systems that exists at the atomic level is now echoed in systems that are subject to gravitation. This has worrying consequences when we consider that spacetime is believed to be substantial – a single substantive spacetime cannot exhibit two distinct forms at the same position (as this interpretation of the flyby anomaly requires). As we will see, a relational model of spacetime is also fraught with problems and appears unsatisfactory.

The question of whether the Hubble expansion absolute or relative to an observer is important. This can be addressed by considering the hypothetical situation represented in Fig. 4.

If bound entity G suddenly becomes unbound, for example through a collision, bodies A – F, observing G, cannot each subsequently register a recession velocity proportional to their distance from G and in the direction pointing away from G as would be expected if the recession were relative. This would lead to an inconsistency – D would eventually see G colliding with B (or even stranger, passing through it), B would see G hitting D, and so on. The expansion cannot therefore be a strictly relative phenomenon when viewed by bound observers. Looking at the situation from the privileged viewpoint of external unbound observers, O and O′ (expanding with respect to one another), G will appear to move away from P, exiting the hexagon by passing between C and D. For consistency then, bound observers must conclude that the expansion taking place with respect to the special point P, the point where the composite entity is linked into the expanding frame. This conclusion holds for any viable model of spacetime (even that proposed by Cooperstock *et al.*).

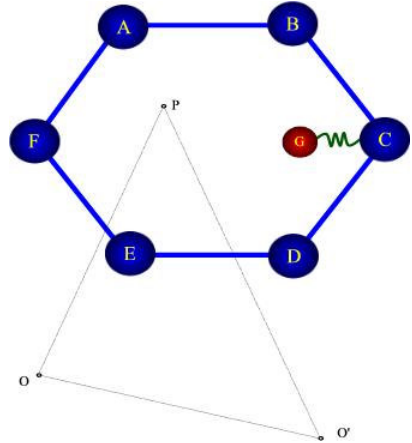

**Figure 4** Entities A – G are bound and are therefore not subject to the cosmological expansion. Unbound observers O and O′ see the complete system moving away them with point P as the intersection of the lines of motion. Assume G becomes instantaneously unbound but does not change location or proper velocity with respect to O and O′. How will G subsequently move (or appear to move) such that the motion is consistent for all bound and unbound observers?

How is this point P selected? Since motion with respect to the expansion is opposed by an acceleration $a_o$, introduced to conserve energy, energy will only be conserved if the force is applied to the centre of mass (otherwise a torque is introduced). The line of the acceleration must coincide with the centre of mass for observers O and O′. P is therefore the centre of mass of the bound system.

This clarifies our understanding of the expansion considerably and happily makes the introduction of multiple expanding spacetimes unnecessary. We have all but exorcised the spectre of differing relative expansions, but for one reasonable enquiry: It is believed the Solar System is bound into the Milky Way galaxy; how then can the Pioneer spacecraft join the expansion? We do not expect an object bound into the galaxy to be able to do so. The only reasonable conclusion is that the Sun is not bound to the galaxy centre. This is possible – we saw earlier the threshold binding energy is not necessarily zero. Is there any evidence to support this radical assertion? It is technically possible to determine whether the Solar System is bound to the galaxy from an examination of the cosmic microwave background radiation (CMBR), but we need to determine if the accuracy of the data currently available from the COBE and WMAP satellites is good enough to enable any conclusions to be drawn.

The cosmic background radiation is expanding with respect to the centre of mass of the system in which the observer is bound. This may be the Sun, the centre of the Galaxy, the centre of the Cluster, or even the centre of the Super Cluster. If this were not the case, isolated observers within the same bound system would disagree about the observations. Let us assume the centre of mass of the system in which the WMAP satellite is located bears the relationship to the satellite shown in Fig. 5, and further assume that the relative velocity of the observation satellite with respect to the centre of mass is zero.

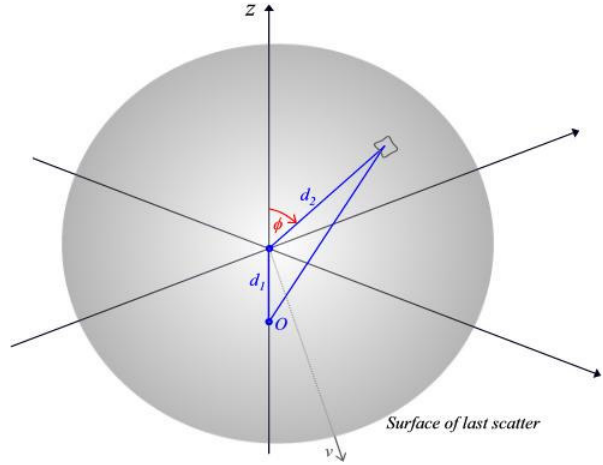

**Figure 5** Define the direction z in the line from the observer O to the centre of mass at the origin. The surface of last scatter is centred on the origin.

The recession velocity can be treated as a proper velocity and the problem dealt with Special Relativistically[‡]. Because of their displacement from the centre of the expanding shell, the observer will register a small difference in the apparent radial velocity that varies with $\phi$. This will affect the redshift and therefore the measured blackbody temperature, but only to a very small extent because the reduction in radial redshift factor is partially compensated for by the transverse redshift that now appears: Let the measured temperature be $T$ and the source temperature $T_o$, the two temperatures being related by the expression $T = T_o / (1+z)$ where $z$ is the redshift factor (believed to be about 1100). Calculating $\Delta(1+z)$, the deviation from the value at $\phi = 0$, slotting this into the relation $\Delta T/T = -\Delta(1+z)/(1+z)$ and applying sensible approximations, the change in apparent temperature as a function of $\phi$ is:

$$\frac{\Delta T}{T} \cong \frac{v}{c} \frac{\gamma}{(1+z)} \frac{d_1^2}{d_2^2} \frac{\sin^2 \phi}{2}. \qquad (3)$$

Recognising that $v/c \sim 1$ whereupon the Lorentz factor $\gamma \sim (1+z)/2$, the perturbation results in the CMBR being relatively colder in the $+z$, $-z$ directions and hotter in the $x$-$y$ plane. Expressed as a wave, the amplitude of the variation is $d_1^2/2d_2^2$. Of course $d_2$ is the distance travelled by the CMBR photons, which is around $1.3 \times 10^{26}$ m. If the Sun is bound to the galaxy centre, $d_1$ is then the distance to the centre of the galaxy, at least $2.5 \times 10^{20}$ m. The change in apparent temperature is seen to be completely insignificant.

However, if the Sun is bound into the galaxy, the redshift change is not the only way the microwave background is affected by a displacement of the observer from the centre of mass. The Sun is at least 26,000 light years away from the centre of the galaxy and one should recognise the surface of last scattering (from where the microwave background originates) is not really a surface at all but possesses significant depth (possibly as much as $\Delta z = 100$). This means that CMBR photons received from the $+z$ direction do not come from the same expanding shell (with respect to the centre) as those in the $-z$ direction. This follows from the time difference that exists between a displaced observer and the centre of the galaxy. A Solar System observer looking in the $+z$ direction will see photons that left 26,000 years earlier in the galaxy frame, or 47 years earlier in the CMBR frame (applying the relativistic time contraction factor). In the opposite direction, photons are received from the surface of last scattering when the surface is 47 years older. This is significant because the temperature of the Universe changes over time, hence observations in the +z and –z directions are not measuring the source at the same temperature. The detailed temperature development in the Universe through the surface of last scattering is discussed by Lachièze-Rey and Gunzig[10], and the consensus model predicts that in that matter-dominated phase of the Universe, the temperature follows the relation:

$$T_o \propto \frac{1}{t^{-2/3}},  \quad (4)$$

where $t$ is the age of the Universe at the time. This also follows from a thermodynamic analysis and consideration of the energy lost through the CMBR photons. The change in temperature over 47 years is found by differentiating relation (4),

$$\frac{\Delta T}{T} = -\frac{2}{3}\frac{\Delta t}{t}.  \quad (5)$$

---

[‡] This may seem strange when the underlying metric is general relativistic, but bear in mind there is some precedence for this: Special Relativity is consistent with the lengthening of supernova light curves and is applied to explain the apparent superluminal motion of some distant objects. In any case, the spacetime through which the photons travel has a curvature of zero.

Taking $t$ as 400,000 years, this effect should introduce a dipole with the sky hotter by 214 µK in the direction of the galaxy and colder by the same amount away from the galaxy. This is shown in Fig. 6 where it is compared with the dipole associated with the proper motion with respect to the CMBR. The dipole associated with the displacement of a bound Solar System from the centre of the galaxy is about 15 times smaller and is not visually evident in the data, but this is not surprising given its smallness. If the dipole does exist, then there are two non-aligned dipoles in the data and these should be removed separately - cleaning up the data by removing a single dipole only will result in the appearance of spurious residual quadrupole and higher order harmonics aligned with the proper velocity dipole and the galactic plane. In fact unexplained quadrupole and octopole components have been found (Fig. 7), but these seem to be related to the ecliptic rather than the galactic plane[11].

We can therefore conclude there is no evidence in the CMBR data the Solar System is bound into the galaxy, thus we may state that it is entirely possible for the Pioneer probes to join the Hubble flow once the barycentric orbital trajectories becomes hyperbolic.

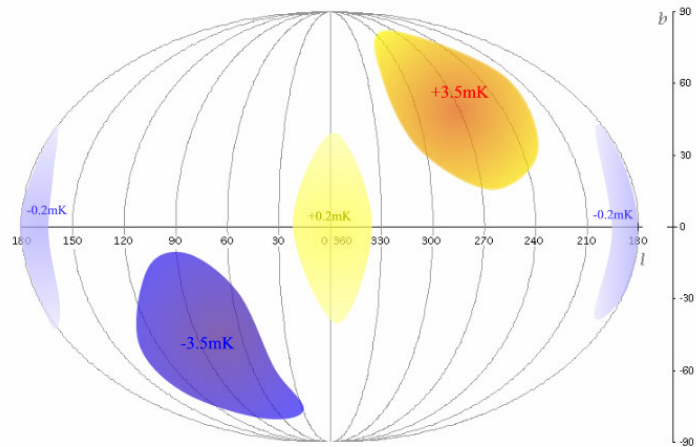

**Figure 6** A representation of the CMBR map showing the proper velocity dipole (+3.5 mK/-3.5 mK) and the much smaller dipole associated with the displacement of the observation platform from the centre of mass (assuming the Sun is bound to the centre of the Milky Way galaxy)

The WMAP satellite is located at the L2 Lagrange point outside Earth orbit and is undoubtedly bound into the Solar System. The separation from the barycentre is too small for there to be any effect on CMBR data through the mechanisms previously shown to be relevant on the scale of the galaxy. This makes the alignment of the observed quadrupole and octopole anomalies with the ecliptic puzzling (Fig. 7). Of course, we have up to now neglected the proper velocity of the observer – the WMAP satellite has a barycentric transverse velocity of around 30 km s$^{-1}$, but its effect is routinely removed by applying a $\boldsymbol{u}.\hat{\boldsymbol{r}}$ correction ($\boldsymbol{u}$ is the proper velocity vector and $\boldsymbol{r}$ the direction of observation).

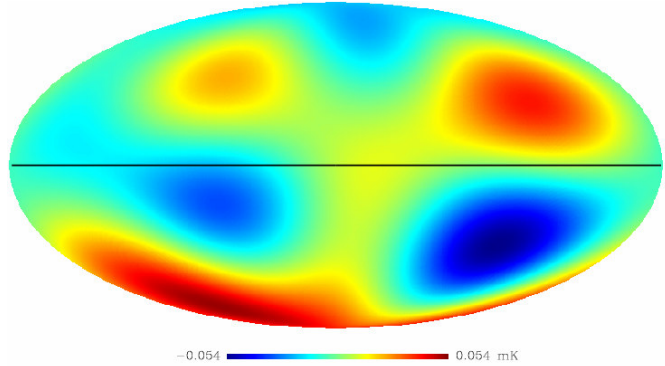

**Figure 7** An ecliptic coordinate representation of the CMBR quadrupole and octopole overlaid to show the unlikely separation of hot and cold spots by the ecliptic plane (black line). There is weak evidence for galactic alignment and stronger evidence for an alignment with the perpendicular plane to the proper motion dipole. The image is reproduced from http://www.phys.cwru.edu/projects/mpvectors/ (accessed 20/10/2006).

Lachièze-Rey and Gunzig[10] calculate this correction simply by applying the non-relativistic Doppler shift associated with a proper velocity. Treating the source velocity also as a proper velocity gives exactly the same result if the relative velocity is found by using the Special Relativistic equation for the addition of velocities: Consider an observer source velocity boost $u$ ($= \Delta u$) away from the direction of observation, with a source recession velocity of $v$. The redshift change is

$$\frac{\Delta(1+z)}{\Delta u} = \frac{\partial}{\partial v}\left(\gamma\left(1+\frac{v}{c}\right)\right)\frac{\partial v}{\partial u}, \qquad (6)$$

and $v + \partial v = (v + \partial u)/(1 + v\partial u/c^2)$, the equation for the relativistic addition of velocities. After differentiation and some trivial algebra, equation (6) simplifies to

$$\frac{\Delta(1+z)}{(1+z)} = \frac{\Delta u}{c}. \qquad (7)$$

This of course is the same as the non-relativistic correction for a proper velocity $v \ll c$. The transverse Doppler effect is not significant in this case, and there is no propagation time lag associated with a velocity change on the part of the observer (this is not entirely obvious, but can be demonstrated by requiring a continuity of interaction through an observer acceleration[12]) hence observations in all directions are with respect to the same emission time. Although the correction applied for the proper velocity appears to be completely correct, we must conclude that still exists an unknown effect has not been accounted for and results in the association with the residuals and the ecliptic plane. One possibility is that the unknown effect may be associated with the motion of an observer with respect to the centre of mass that appears as an additional redshift on unbound sources. There is possibly some support for this in the sinusoidal variation in the Doppler residuals recorded for the NEAR flyby and also the annual and diurnal oscillations on the Pioneer data.

VI Discussion and Further Work

Though the approach has been rather *ad hoc*, the possibility that the transition from a bound to an unbound orbit is associated with an entity joining the Hubble flow has some qualitative and quantitative support. Nevertheless the concept is problematic because the need for two distinct spacetimes is irreconcilable with the generally accepted substantive model of spacetime, and it is very difficult to understand how boundedness could be locally determined (a necessity in an alternative relationist interpretation of spacetime). A further problem is the infinity occurring in the energy relationship postulated between the bound and unbound states.

In spite of the problems (which are largely conceptual and one should bear in mind that standard physics and cosmology has only been stringently tested over what is a very restricted local domain in comparison to the full extent of the Universe), the idea is worth pursuing because of the possibility of understanding the rather surprising link between the CMBR quadrupole and octopole components and the ecliptic. The effect of the velocity and acceleration of the observer with respect to the bound centre of mass on observed photons from objects in the Hubble flow has not yet been explored, and might offer a simple way of resolving the anomaly without having to invoke improbable geometries and/or hidden matter. It has been shown that in the basic linear expression for the energy relationship between bound and unbound states, the displacement alone does not have a significant effect on the scale of the Solar System, but there is no obvious additional velocity or acceleration effects emerging from a Special Relativistically consistent treatment. It is therefore probable that a more sophisticated formulation of the energy relationship is required[§]. There may be clues as to how this should be developed by examining the annual and diurnal periodicities in the Pioneer Doppler residuals and matching these to the velocity and acceleration of the Earth, though it should be born in mind that the periodicities are interpreted by Anderson *et al.*[1] as most likely a systematic associated with the modelling of the Earth's orientation and the accuracy of the planetary ephemeris. However, the annual effect has a turning point exactly at opposition and this coincidence is suggestive of a real effect. Taking a heuristic approach, the annual oscillation could be associated either with the radial velocity (zero at opposition) or the radial acceleration (maximum at opposition) – all other possibilities are ruled out through symmetry considerations. If the annual oscillation is an additional acceleration on the spacecraft, the association is with the radial velocity of the Earth; on the other hand, it may be an additional unmodelled Doppler shift directly associated with the radial acceleration. Accurate ranging data is needed differentiate.

It may also be possible to determine the relationship from an analysis of the NEAR EGA (Fig. 2), because both Doppler and range data are available and the positional relationship between the observer and the spacecraft is accurately known.

---

[§] Or it may be a frequency change associated with a complex three-body effect (observer –centre of mass- source)